\documentclass[aps,prl,twocolumn,preprintnumbers,superscriptaddress,nobibnotes,floatfix,longbibliography]{revtex4-1}

\pdfoutput=1
\usepackage{amsmath,amsfonts,amssymb,mathrsfs,graphicx,color,longtable}
\usepackage{hyperref}
\usepackage{graphicx}
\usepackage{amsfonts,amsmath,amssymb,bm}
\usepackage{array}
\usepackage{color}
\usepackage{enumitem}
\usepackage[explicit]{titlesec}
\usepackage[utf8]{inputenc}

\hypersetup{
     colorlinks   = true,
     citecolor    = blue,
     urlcolor     = blue,
     linkcolor    = blue
}

\usepackage{xcolor}

\newcommand{\MDM}{M_{\text{DM}}}

\newcommand{\be}{\begin{equation}}
\newcommand{\ee}{\end{equation}}

\newcommand{\PRL}{Phys. Rev. Lett.}

\graphicspath{{figs/}}

\makeatletter

\def\hhref#1{\href{http://arxiv.org/abs/#1}{arXiv:#1}}
\usepackage{xstring} 
\newcommand{\hhrefq}[1]{\IfSubStr{#1}{:}{\href{http://inspirehep.net/search?ln=en&ln=en&p=#1&of=hb&action_search=Search&sf=&so=d&rm=&rg=25&sc=0}{InSpires:#1}}{\hhref{#1}}}

\def\art{\@ifnextchar[{\eart}{\oart}}
\def\eart[#1]#2#3#4#5#6{{\rm #2}, {\em #3 \bf #4} {\rm (#6) #5} ({\em #1})}
\def\article{\@ifnextchar[{\earticle}{\oarticle}}
\def\oarticle#1#2#3#4#5#6{{\rm #1}, {``#6''}, {\rm #2 #3 (#5) #4}}
\def\earticle[#1]#2#3#4#5#6#7{{\rm #2}, {``#7''}, {\rm #3 #4 (#6) #5}  [\hhrefq{#1}]}
\def\hepart[#1]#2{{\rm #2, \sl#1}}
\def\heparticle[#1]#2#3{#2, { ``#3''} [\hhrefq{#1}]}

\begin{document}

\title{Co-SIMP Miracle}

\author{Juri Smirnov}
\thanks{{\scriptsize Email}: \href{mailto:smirnov.9@osu.edu}{smirnov.9@osu.edu}; {\scriptsize ORCID}: \href{http://orcid.org/0000-0002-3082-0929}{ 0000-0002-3082-0929}}
\affiliation{Center for Cosmology and AstroParticle Physics (CCAPP), The Ohio State University, Columbus, OH 43210, USA}
\affiliation{Department of Physics, The Ohio State University, Columbus, OH 43210, USA}

\author{John F. Beacom }
\thanks{{\scriptsize Email}: \href{mailto:beacom.7@osu.edu}{beacom.7@osu.edu}; {\scriptsize ORCID}: \href{http://orcid.org/0000-0002-0005-2631}{0000-0002-0005-2631}}
\affiliation{Center for Cosmology and AstroParticle Physics (CCAPP), The Ohio State University, Columbus, OH 43210, USA}
\affiliation{Department of Physics, The Ohio State University, Columbus, OH 43210, USA}
\affiliation{Department of Astronomy, The Ohio State University, Columbus, OH 43210, USA}

\date{10 February 2020; updated 26 June 2020}

\begin{abstract}
We present a new mechanism for thermally produced dark matter, based on a semi-annihilation-like process, $\chi+  \chi +\text{SM} \rightarrow \chi  + \text{SM}$, with intriguing consequences for the properties of dark matter.  First, its mass is low, $\lesssim 1$ GeV (but $\gtrsim 5$ keV to avoid structure-formation constraints).  Second, it is strongly interacting, leading to kinetic equilibrium between the dark and visible sectors, avoiding the structure-formation problems of $\chi+  \chi + \chi  \rightarrow \chi + \chi$ models.  Third, in the $3 \rightarrow 2$ process, one dark matter particle is consumed, giving the standard-model particle a monoenergetic recoil.  We show that this new scenario is presently allowed, which is surprising (perhaps a ``minor miracle").  However, it can be systematically tested by novel analyses in present and near-term experiments. 
In particular, the Co-SIMP model for thermal-relic dark matter can explain the XENON1T excess.
\end{abstract}  

\maketitle


{\bf Introduction.---}
\label{sec:intro}
For dark matter (DM) models, thermal production mechanisms are highly predictive frameworks~\cite{Kolb:1990vq, hep-ph/9506380, 0404175}.  Charting possible realizations is important, as it leads to insights that guide experimental efforts to fully test particle dark matter.  A thermal production process, if confirmed experimentally, would provide a new probe of the physical conditions of the early universe.

The best studied thermal candidate is the WIMP (weakly interacting massive particle)~\cite{Zeldovich:1965gev, Lee:1977ua, Steigman:1984ac, Arcadi:2017kky, Roszkowski:2017nbc}.  In the simplest case, the annihilation cross section to all final states is determined from the relic abundance as $\langle \sigma v \rangle  = (2.2 \times 10^{-26} \, \text{cm}^3 \, \text{s}^{-1}) (0.12 / \Omega_{\rm DM} h^2)$~\cite{SteigmanDasguptaBeacom}. The ``WIMP window" is defined by the smallest mass allowed by annihilation constraints (20 GeV if neutrinos are neglected~\cite{1805.10305}; 10 MeV if they are dominant~\cite{1411.6005}) and the largest mass allowed by unitarity (150 TeV~\cite{GriestKamionkowski,1904.11503}).

It is important to consider other possibilities~\cite{1705.03689,1003.5912, 1607.03108, 1702.07716, SIMP, 1612.09074,1812.11418,1907.08262,1803.02901,1906.00981,1906.09269,SIMPlest}. Recent work~\cite{SIMP, SIMPlest} has made the simple but ingenious point that the process $\chi + \chi + \chi \rightarrow \chi + \chi$ is efficient in the early universe if the interactions are strong, setting the relic abundance while involving only dark-matter processes, hence the name strongly interacting massive particle (SIMP). The observed DM abundance requires $\langle \sigma_{32}v_{\rm rel}^2 \rangle  M_{\rm DM}^2 \approx 10^8 \text{ GeV}^{-3}$~\cite{ThermalHistory,SIMP}.  Assuming a scaling behavior of $ \langle \sigma_{32}v_{\rm rel}^2 \rangle \equiv \alpha_{\rm eff}^3/M_{\rm DM}^5$ implies an MeV-scale dark matter mass~\cite{SIMP}. A dark matter sector that converts DM rest mass into kinetic energy that is kinetically decoupled from the standard model (SM) will heat itself up~\cite{Schmalz}. The DM free-streaming length would then be too long, as the DM particles would be too fast. To dissipate the heat into the SM sector and to slow down the DM particles, an elastic SM-SIMP interaction, $\chi + \text{SM} \rightarrow \chi + \text{SM}$, has to be postulated~\cite{SIMP}.  However, in the mass range relevant for the SIMPs, it is hard to do that without inducing new $\chi + \chi  \rightarrow \chi + \chi $ interactions~\cite{SIMPlest, NNLOSIMPs} that conflict with cluster observations~\cite{BulletCluster,1504.03388,1504.06576,1608.08630} (though there may be ways out~\cite{1806.00609,1504.00745,1505.00960,1601.03566,1801.07726}). 

\begin{figure}[t]
\includegraphics[width=0.35\textwidth]{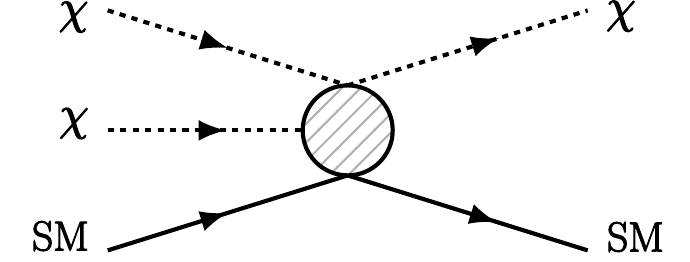}
\caption{The Co-SIMP freezeout process, which also keeps DM in kinetic equilibrium with the SM.  This process gives novel monoenergetic-recoil signals in detectors.}
\label{fig:CoSIMP}
\end{figure}

We take a different approach and suppose that the DM decouples through $\chi  + \chi + \text{SM} \rightarrow \chi + \text{SM}$.  Figure~\ref{fig:CoSIMP} shows this number-changing interaction for the DM, which also keeps it in kinetic equilibrium with the SM plasma, avoiding overheating the DM.  As detailed below, what we call the Co-SIMP mechanism leads to a dark-matter candidate with vastly different properties and phenomenology from other thermal relics.  Despite the Co-SIMP's low mass and strong interactions, it seems to have evaded all present constraints. However, testing Co-SIMPs is within reach.

In the following, we define the Co-SIMP model, calculate the corresponding freezeout process and its consequences, then predict present-day signals based on Fig.~\ref{fig:CoSIMP}, as well as signals expected from loop-induced elastic interactions, and conclude.


{\bf Co-SIMP Interactions.---}
Given the Co-SIMP interaction in Fig.~\ref{fig:CoSIMP}, the dark sector must have a $Z_3$ symmetry (this could be generalized to $Z_N$ with $N >3$) to ensure DM stability, similarly to Ref.~\cite{1003.5912}. To prevent a WIMP-like freezeout ($\chi + \chi \rightarrow   \chi+ \text{SM} + \text{SM}$), we require $M_{\rm DM} \leqslant 2 M_{\rm SM}$.  Direct couplings to photons or neutrinos are prohibited to avoid $2 \rightarrow 3$ processes such as $ \chi + \chi \rightarrow  \chi + \gamma + \gamma$.  The final state of the freezeout process is, for $\MDM \ll M_{\rm SM}$, a semi-relativistic $\chi$ and a non-relativisitic SM particle. For the lower mass bound, we take $\MDM > 5 \,\rm keV$ to avoid structure-formation constraints~\cite{1908.06983, 1912.09397}.  More specific constraints are discussed below.

The Co-SIMP upper mass bound varies depending on the DM interaction operator. For a leptophilic model, coupled via $\mathcal{O}_\ell =  \bar{\ell}_\alpha \ell_\beta \chi^3/\Lambda^2$, we require $M_{\rm DM} \lesssim \rm MeV$. For a nucleophilic model coupled via  $\mathcal{O}_n =  \bar{n} n \chi^3/\Lambda^2$, $\mathcal{O}_p =  \bar{p} p \chi^3/\Lambda^2$, or $\mathcal{O}_\pi = \pi^2  \chi^3/\Lambda$, we require $M_{\rm DM} \lesssim \rm GeV$. We focus on the leptophilic and nucleophilic scenarios. (A more exotic possibility arises if coupling to weak gauge bosons via $\mathcal{O}_W =  F_{\mu \nu}^W F^{W \, \mu \nu } \chi^3/\Lambda^3$, or top-quarks only, is allowed, in which case the DM mass can be as large as $\sim 100$ GeV.  In the case of coupling to $W$ bosons, a particular construction of the UV-completion is required in order to avoid direct photon couplings.)

Our focus is on defining a new framework for thermal DM; see the Supplemental Material for further exploration of the details, including possible UV completions, which are crucial for investigating Co-SIMP interactions at colliders and other high energy environments.

{\bf Co-SIMP Freezeout.---}
The Boltzmann equation for freezeout via $\chi  + \chi + \text{SM} \rightarrow \chi + \text{SM}$ is
\begin{align}
& s H(T) z \, Y'_{\rm DM} = - \gamma_{32} \left( \frac{n_{\rm DM}^2}{n_{\rm DM, eq}^2} - \frac{n_{\rm DM}}{n_{\rm DM, eq}}\right)  \nonumber \\
& = - \langle \sigma_{32}v_{\rm rel}^2 \rangle \left( n_{\rm DM}^2 - n_{\rm DM} n_{\rm DM, eq} \right) n_{\rm SM, eq}\,.
\end{align}
The space-time interaction density is approximated as $\gamma_{32}  \approx  \langle \sigma_{32}v_{\rm rel}^2 \rangle  \, n_{\rm DM, eq}^2\,n_{\rm SM, eq}$ in the non-relativistic regime. The Hubble rate is $H(T)$ and $s$ the entropy density. 
Defining $z = \MDM/T$ and $Y'_{\rm DM} = dY_{\rm DM}/dz$, we get
\begin{align}
Y'_{\rm DM} = - \frac{\lambda_{32}}{z^5} \left( Y_{\rm DM}^2 - Y_{\rm DM} \,Y_{\rm DM, eq}\right).
\end{align}
The dimensionless quantities are 
\begin{equation}
\lambda_{\rm 32}
= \left(  Y_{\rm SM}^{\rm sym.}+ \eta\right)\times \frac{ \langle \sigma_{32}v_{\rm rel}^2 \rangle \,s^2}{H(T) }\biggr\rvert_{M_{\rm DM}} \,,
\end{equation}
with $Y_{\rm DM} = n_{\rm DM}/s$, $Y^{\rm sym.}_{\rm SM} = n^{\rm sym.}_{\rm SM}/s$ being the symmetric SM particle abundance, and $\eta$ the baryon-to-photon ratio. Application of the boundary-layer method~\cite{bender, ThermalHistory} yields the asymptotic relic abundance 
\begin{align}
Y_{\rm DM}(\infty) \approx - \left( \int_{z_{\rm f.o.}}^{\infty} \frac{\lambda_{\rm 32}}{z^5} dz \right)^{-1}\,.
\end{align}
The freezeout temperature is defined by $z_{\rm f.o.} \approx \log{\lambda_{32}}- 5/2 \log{z_{\rm f.o.}} -1.65$, which gives $z_{\rm f.o.} \approx 10$.  Note the dependence of the interaction factor on $v^2_{\rm rel.}$, which results from an incoming flux of two particles on target. At leading order, in exothermic processes $\sigma_{32} v^2_{\rm rel.} \sim \rm const.$~\cite{1702.07716}. 

The relic density is $\Omega_{\rm DM} \approx Y_{\rm DM}(\infty) M_{\rm DM} s_0/\rho_{\rm crit.}$.  For the typical case, $\MDM \ll M_{\rm SM}$, the cross section of the number-changing interaction is 
\begin{align}
 \langle \sigma_{32}v_{\rm rel}^2 \rangle_{\rm f.o.}
\approx 10^{12}\, \left(\frac{10^{-9}}{\eta} \right) \left(\frac{ \text{MeV}}{M_{\rm DM}} \right)^{3}\left(\frac{0.12}{ \Omega_{\rm DM} h^2} \right) \text{GeV}^{-5} \,.
\end{align}
For the edge case, $\MDM \approx M_{\rm SM}$, the decoupling of Co-SIMP interactions takes place at higher temperatures, such that the Co-SIMP freezeout happens dominantly through interactions with the symmetric component of the SM particle bath, leading to
\begin{align}
 \langle \sigma_{32}v_{\rm rel}^2 \rangle_{\rm f.o.}
\approx 5\times 10^{3}\,  \left(\frac{10 \, \text{MeV}}{M_{\rm DM}} \right)^{3}\left(\frac{0.12}{ \Omega_{\rm DM} h^2} \right) \text{GeV}^{-5} \,.
\end{align}
These are the central predictions of the new Co-SIMP thermal production mechanism.
The loop-induced process $\chi \chi \rightarrow \chi + 2 \gamma$ is subdominant during the freezeout process throughout the Co-SIMP parameter space.  Late-universe annihilations that produce gamma rays are discussed in the Supplemental Material.

Considerations related to light-element dissociation during Big-Bang Nucleosynthesis (BBN)~\cite{1808.09324,1809.01179,1904.04256} do not constrain Co-SIMPs, because the energy released into SM particles is much smaller than for WIMP freezeout.

However, in the leptophilic case, because the Co-SIMP can have $\MDM \ll m_e$, the number of relativistic degrees of freedom $N_{\rm eff}$ is affected, in fact being reduced from the standard value.  We discuss this in detail in the Supplemental Material, considering both BBN and Cosmic Microwave Background (CMB) observations.
Because of this negative contribution, if experiments find a Co-SIMP mass well below the electron mass, it would be an indication of a complex early history. For example, a cancellation with a positive contribution to $\Delta N_{\rm eff}$ could take place to explain the current data, which are consistent with $\Delta N_{\rm eff} \approx 0$~\cite{1807.06209}. 

In the nucleophilic case, the Co-SIMP masses are above the MeV scale, so that BBN observables and $\Delta N_{\rm eff}$ are not affected.


{\bf Testing the Co-SIMP Process.---}
Strikingly, the Co-SIMP freezeout process ($\chi + \chi + \text{SM} \rightarrow \chi + \text{SM}$) happens directly in detectors  with a significant rate, unlike for WIMPs ($\chi + \chi \rightarrow \text{SM}+\text{SM}$)~\cite{hep-ph/9403357}. This is possible because of the strong DM interactions, the high DM density due to the low DM mass, and the high SM density of ordinary materials.

The $3 \rightarrow 2$ process produces energetic SM particles in a detector, like WIMP direct detection, but with important differences.  First, the kinetic energy of the SM particle, provided by the consumption of one DM particle,  is monoenergetic. This follows directly from the fact that in the CM frame, the outgoing particles have equal momenta. The SM recoil energy is 
\begin{equation}
E_R \approx \frac{3 \MDM^2 \left(  \MDM + 2 M_{\rm SM} \right) \left( 3  \MDM + 2 M_{\rm SM} \right) }{8 M_{\rm SM}  \left( 2 \MDM + M_{\rm SM} \right)^2}\,,
\label{eq:recoilenergy}
\end{equation}
which reduces to $E_R \approx 3/2 \, \MDM^2/M_{\rm SM}$ for the typical case, $\MDM \ll M_{\rm SM}$, and to $E_R \approx 5/8 \, \MDM$ for the edge case, $\MDM \approx 2 M_{\rm SM}$.  The kinetic energy of the incoming DM particle is negligible in this reaction. Second, though the kinetic energy of the SM particle will be at most barely relativistic, it can be well above the energy produced by a recoiling WIMP of comparable mass. For the WIMP case, the recoil energy is  $E_R \approx 2 v^2 \mu^2/M_{\rm SM}$, with $\mu$ being the reduced mass~\cite{1903.03026}, which is a factor $v^2\sim 10^{-6}$ smaller than for the Co-SIMP case. Those two points make for a signal that could stand out from backgrounds. (Other scenarios with large energy deposits involve DM de-excitation~\cite{1707.05380,1712.00455,1905.12635,1908.10861,2001.11514}.) Third, even if elastic scatterings in the detector overburden slow the Co-SIMPs down, the expected signal spectrum shape is unaltered because it depends on the loss of DM rest mass, not kinetic energy.

The event rate per detector volume is
\begin{align}
& \gamma_{32}= \frac{R}{V} \approx   \langle \sigma_{32} v_{\rm rel}^2 \rangle n^2_{\rm DM} n_{\rm SM}  \\ \nonumber
& \approx  \frac{0.8}{\text{m}^3 \text{ day} } \left( \frac{n_{\rm SM}}{ N_A \,\text{cm}^{-3} } \right) \left( \frac{0.1\, \rm MeV}{M_{\rm DM}}\right)^2 \left( \frac{ \langle \sigma_{32} v_{\rm rel}^2 \rangle  }{10^{16} \text{ GeV}^{-5}}\right)\,, \nonumber
\end{align}
where $n_{\rm SM}$ and $n_{\rm DM}$ are the SM and DM number densities, $N_A$ is the Avogadro number, and we use the observed DM mass density.  

\begin{figure}[t]
\includegraphics[width=0.49\textwidth]{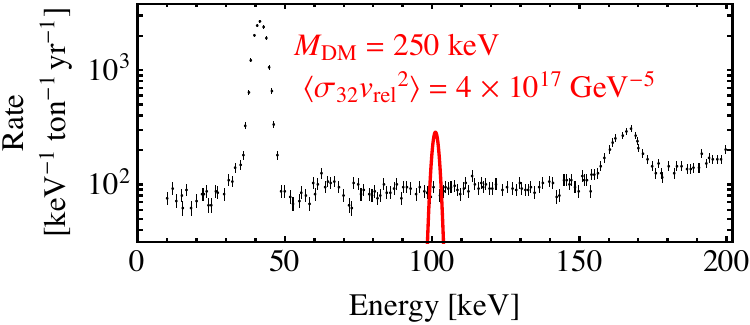}
\caption{Monoenergetic electron recoil spectrum caused by the Co-SIMP process for $\MDM = 250 \text{ keV}$ (hence $E_R \approx 101 \text{ keV}$), including energy resolution, compared to XENON1T data.  For better visibility, $\langle \sigma_{32} v_{\rm rel.}^2\rangle$ is chosen so that the signal is ten times greater than needed for a $90\%$ CL exclusion ($1.28\sigma$).}
\label{fig:DDSpectrum}
\end{figure}

Figure~\ref{fig:DDSpectrum} shows the monoenergetic spectrum in XENON1T caused by the leptophilic Co-SIMP process.  We include energy resolution, which is $3.5\%$ in this energy range~\cite{1904.11002}.  Superposed are their measured electron recoil data. This shows that a bump hunt could be highly efficient for testing the Co-SIMP scenario. 

\begin{figure}[t]
\includegraphics[width=0.48\textwidth]{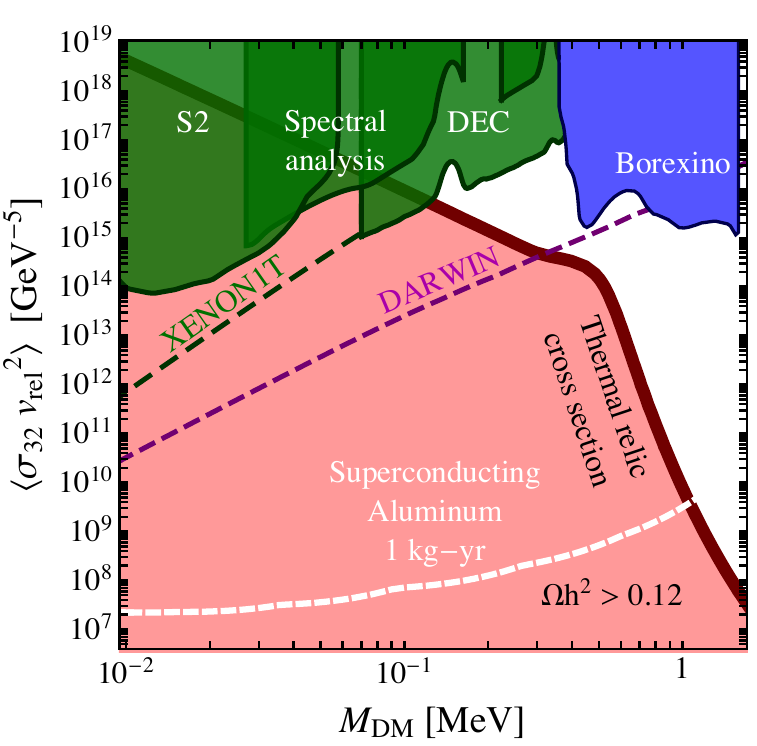}
\caption{Current bounds and projected sensitivities for electrophilic Co-SIMPs. The parameter space is largely open, despite the low DM mass and strong couplings. The relic density at $\MDM \ll m_e$ is affected by the baryon asymmetry, leading to a change of the slope. We show bounds based on the Co-SIMP process in detectors (colored regions), along with projected sensitivities (dashed lines). A low-threshold detector based on superconducting aluminum could test nearly all the parameter space (white dashed line). In the pink region, the relic density is too high; at its boundary, it is correct.
}
\label{fig:DDconstraint}
\end{figure}

Figure~\ref{fig:DDconstraint} shows the Co-SIMP parameter space, including current constraints.
For the Borexino experiment~\cite{hep-ex/0012030}, we convolve the predicted signal with the energy resolution and compare it to the measured data in Ref.~\cite{1707.09279}. 
The uncertainty scale is set by the square root of the number of measured events in a bin of width somewhat larger than the energy resolution. 
This is appropriate because the backgrounds are well modeled.  For XENON1T, we have an analogous procedure.  We use data from their double electron capture (DEC) search~\cite{1904.11002} and their S2-only light DM search~\cite{1907.11485}.  At intermediate energies, we use data from their electron recoil study~\cite{baudis}, conservatively requiring that the signal be less than the measured events in a bin because the background analysis is still preliminary. (In the Supplemental Material, we repeat the analysis taking into account the xenon orbital effects, this weakens the bounds somewhat for the low Co-SIMP masses.)

Figure~\ref{fig:DDconstraint} also shows the projected sensitivity for 1 ton-yr of XENON1T, assuming a dedicated line search, with well-modeled backgrounds, in their entire electron recoil energy range. Similarly, we show the sensitivity of the proposed DARWIN experiment in $200$ ton-yr~\cite{1606.07001}. Other relevant experiments to search for the Co-SIMP process are KamLAND~\cite{1409.0077} and JUNO~\cite{1507.05613}.

\begin{figure}[t]
\includegraphics[width=0.48\textwidth]{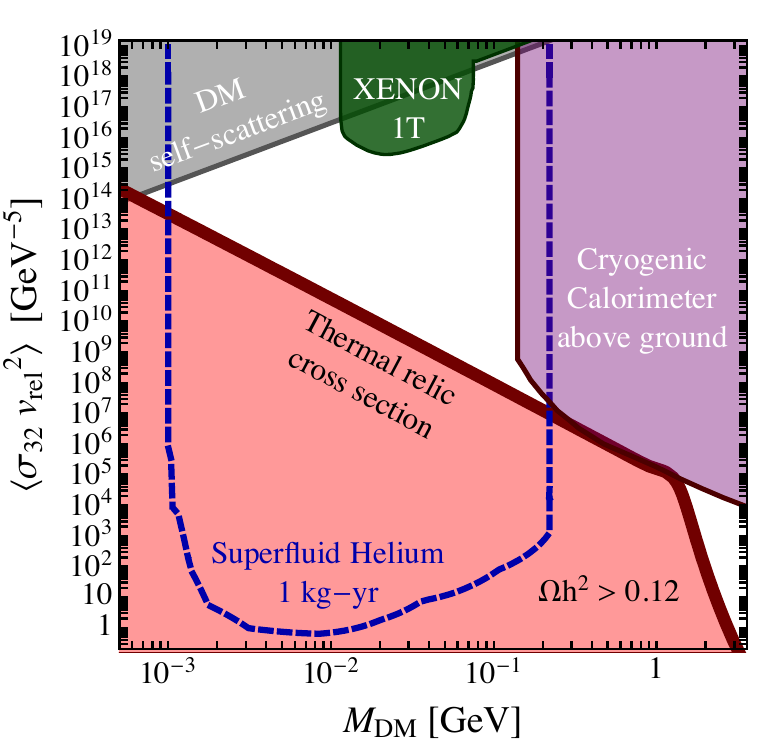}
\caption{Current bounds and projected sensitivities for nucleophilic Co-SIMPs. The change in the relic abundance slope is now related to the proton mass. Limits from the XENON1T experiment constrain the Co-SIMP process directly (dark green region). A detector based on superfluid helium could test a large fraction of the currently viable parameter space (blue dashed line).}
\label{fig:constraintFuture} 
\end{figure}

Figure~\ref{fig:constraintFuture} shows the parameter space for the nucleophilic scenario. It is challenging to test the nucleophilic Co-SIMP process directly in large volume detectors due to lower event rates at larger DM masses. However, loop-induced elastic interactions, discussed below, are efficient for testing nucleophilic interactions. 

\begin{figure}[b]
\includegraphics[width=0.4\textwidth]{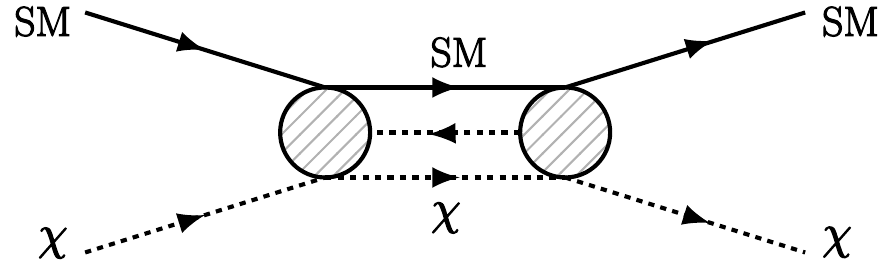}
\caption{Elastic dark matter scattering process with SM particles, induced at two-loop level.}
\label{fig:DD}
\end{figure}

{\bf Elastic Interactions.---}
Figure~\ref{fig:DD} shows that an elastic Co-SIMP scattering process is induced at the two-loop level: $ \text{SM} +    \chi \rightarrow \text{SM}  + \chi$. Given, for example, the electrophilic $\mathcal{O}_\ell$ interaction operator, the induced coupling coefficient $c_{\rm e}^d$ of the interaction $\chi \chi \bar{e } e $ can be computed. After performing the top loop integral, one obtains the following expression (with $ x = \sqrt{1 - 4 M_{\rm DM}^2/k^2}$), which is regularized with a cut-off at the scale $\Lambda$ and analytically approximated, with $M_{\rm SM}$ being the mass of the propagating electron:
\begin{align}
& c_{\rm e}^d = \frac{M_{\rm SM}}{(4 \pi)^2\, \Lambda^4} \int \frac{d^4 k}{(2 \pi)^4}   \log{ \left(  \frac{x +1 }{x -1} \right) } \left( \frac{x }{k^2 - M_{\rm SM}^2}  \right) \nonumber \\ 
&  \approx  \frac{M_{\rm SM}}{(4 \pi)^4 \, \Lambda^2} \left( 1 -  \frac{M_{\rm DM}^2}{\Lambda^2} \right) \log \left( \frac{\Lambda^2 + M_{\rm SM}^2}{4 M_{\rm DM}^2} \right)\,.
\end{align}   
%

To make connection to the freezeout process, we compute the $3 \rightarrow 2$ cross section, given the electron Co-SIMP coupling operator $\mathcal{O}_\ell$, following Ref.~\cite{1702.07716}, and obtain $\langle \sigma_{32} v_{\rm rel.}^2\rangle  \approx \sqrt{3}/(4 \pi \MDM \Lambda^4)$. We can now express the loop coefficient as a function of the $3 \rightarrow 2$ cross section, which yields $( c_{\rm e}^d \, \text{GeV}) \approx 5 \times10^{-9} \sqrt{ \MDM \langle \sigma_{32} v_{\rm rel.}^2\rangle  \text{GeV}^4}$. This leads to DM-electron scattering cross sections of $\sigma_{\rm SI}^{\rm el.} \approx 10^{-37}\,\text{cm}^2$ at $\MDM \sim 100 \, \rm keV$ and $\sigma_{\rm SI}^{\rm el.} \approx 10^{-40}\,\text{cm}^2$ at $\MDM \sim \rm MeV$. Those cross section values are currently unconstrained by direct detection experiments, but are within the reach of future experimental efforts~\cite{1901.07569,1708.06594,1708.08929}.

Figure~\ref{fig:DDconstraint} shows the expected sensitivity of a low threshold detector to the electrophilic scenario. An efficient technology seems to be detectors based on superconducting materials, for example, aluminum~\cite{1512.04533,1709.07882} (here with 1 kg-yr). Two other technologies for direct detection experiments with a low energy threshold~\cite{1708.06594,1708.08929} are discussed in the Supplemental Material.

Figure~\ref{fig:constraintFuture} shows current constraints on the nucleophilic Co-SIMP scenario from a gram-scale cryogenic calorimeter experiment~\cite{1707.06749}. In addition, we find that a detector based on superfluid helium~\cite{1408.3581,1604.08206,1810.06283}, could test a significant fraction of the open parameter space, (here with 1 kg-yr and a threshold of $\sim \rm  meV$). 
For an EFT approach to superfluid helium DM sensitivity see Refs~\cite{1902.02361,1907.10635,1911.04511}. Other techniques, based on polar materials~\cite{1712.06598,1807.10291} could provide comparable sensitivity to nucleophilic Co-SIMPs; see Ref.~\cite{1910.10716} for an overview of possible targets. The loop-induced elastic cross section is computed analogously to the leptophilic case, but where the larger mass increases the numerator. 

{\bf Other Possible Searches.---}
In the Supplemental Material, we discuss a variety of other constraints on Co-SIMPs, none of which are yet as strong and secure as those above. The Co-SIMP process and loop-induced processes can produce x-rays, but we find that the current sensitivities of x-ray satellites~\cite{astro-ph/0102151,1609.00667} are insufficient to be competitive with direct-detection searches. However, future missions, such as Athena~\cite{1807.06092}, will test relevant parameter space.  A variant of the Co-SIMP process, $e^+ + e^- \rightarrow \chi + \chi + \chi$, could cool supernovae and pre-supernova stars~\cite{SNModel,SNrefs}, but its effectiveness is limited by an accompanying opacity due to $\chi + e \rightarrow  \chi + \chi + e$. We estimate the Co-SIMP self-scattering elastic cross section induced at two loops, but we find that it is low in the majority of the parameter space and consistent with observations~\cite{BulletCluster,1504.03388,1504.06576,1608.08630}. (However, in the nucleophilic scenario, it seems to provide the strongest bound on Co-SIMPs with low masses. ) Finally, spectroscopy of SM bound states~\cite{1310.6923} could test the Co-SIMP. Particularly, true muonium~\cite{0904.2225,1904.08458}, with a very compact wave function, would be ideal to test the leptophilic scenario~\cite{1401.6457}, once experimental observations can be performed.  
 
 Elsewhere in this {\it Letter}, we focus on the most model-independent tests of the Co-SIMP model, i.e., those most closely connected to the Co-SIMP production mechanism and at comparable or lower $\sqrt{s}$ to freezeout.  In future work, it would be interesting to explore constraints from higher-energy interactions, though this would require first specifying the UV completion of our framework.  

A key question is if or how Co-SIMPs are constrained by collider missing-energy searches, which are generally quite powerful~\cite{1103.0240,1108.1196,1109.4398}.  Those searches will be ineffective if the produced Co-SIMPs interact with cross sections larger than $\sigma \sim 10^{-30}\, \rm cm^2$ and do not lead to a missing energy signature~\cite{1503.05505}. Thus, we estimate the semi-elastic scattering cross section $\sigma_{\chi + e \rightarrow  \chi + \chi + e} \approx \MDM \sqrt{s}/(256 \pi^3 \Lambda^4) \approx \sqrt{s} \, \MDM^2 \langle \sigma_{32} v_{\rm rel.}^2\rangle/(64 \pi^2 \sqrt{3})$, where we used the previously derived relation between the effective operator scale and the interaction cross section.  For  $\MDM \ll m_e$ this leads to $\sigma_{\chi + e \rightarrow \chi + \chi +e} \approx 5 \times 10^{-26} \sqrt{E_\chi/(100 \, \text{GeV})} \,\text{cm}^2$ and for $\MDM \approx m_e$ to $\sigma_{\chi + e \rightarrow \chi + \chi + e} \approx 10^{-28} \sqrt{E_\chi/(100 \, \text{GeV})}\,\text{cm}^2$. For the nucleophilic case, the cross sections are about an order of magnitude larger since $\sqrt{s}$ is bigger. Thus, even if Co-SIMPs are copiously produced, as expected, existing bounds do not apply, and new analyses would be needed.

A related question concerns cosmic-ray interactions with DM, as in Refs.~\cite{1810.07705,1810.10543,1811.00520,1906.11283,1907.03782}, where large cross sections for light dark matter were ruled out, at least under the assumption of energy-independent elastic interactions (see Ref.~\cite{1907.03782} for a model with energy dependence).  Constraints for Co-SIMPs will depend on the UV completion, and are reserved for future work.


{\bf Conclusions.---} 
It is of high importance to point out thermal production mechanisms of DM, since they lead to highly predictive models and could provide additional information about the conditions of the early universe. We present a new possibility to thermally produce DM. The new ingredient is that the production mechanism predicts the cross section for a process that consumes one DM particle and converts its rest mass into kinetic energy of the catalyzing SM particle.

This mechanism leads to novel signatures, such as through the Co-SIMP freezeout process occurring in detectors, including large ones for neutrinos. At loop level, an elastic scattering process is induced that can be tested in conventional dark matter detectors. Large fractions of the parameter space remain still untested but seem within reach of future dark matter searches, especially through expected sensitivity improvements in the coming years. 


\vspace{0.7cm}
\centerline{\bf Acknowledgments}
\vspace{0.2cm}
We are grateful for helpful discussions with Nicole Bell, Eric Braaten, Christopher Cappiello, Basudeb Dasgupta, Jared Evans, Yonit Hochberg, Francesco Sannino, Bei Zhou, Jure Zupan, Kathryn Zurek, and especially Eric Kuflik and Teresa Marrodan Undagoitia. J.S. is largely supported by a Feodor Lynen Fellowship from the Alexander von Humboldt foundation. The work of J.F.B. was supported by NSF grant PHY-1714479.


\clearpage

\appendix 

\onecolumngrid
\begin{center}
 \bf \large Supplemental Material
\end{center}
\vspace*{0.2cm}

\twocolumngrid

Here we discuss in more detail the effects of light Co-SIMPs on $N_{\rm eff}$, the compatibility of the Co-SIMP scenario with the recently observed XENON1T electron-recoil excess, and other potential searches for Co-SIMPs. None of the latter are yet sensitive enough to test thermally produced Co-SIMPs. However, some of these are likely to become relevant in the near future, due to upcoming or proposed experiments. We end this section with notes on possible UV completions for the Co-SIMP production process.


\subsection{Co-SIMPs and $\Delta N_{\rm eff}$}
The Co-SIMP contribution to the effective number of degrees of freedom in the early universe can be found using conservation of the co-moving entropy,
\begin{align}
\frac{s_\nu(T_{\rm dec})}{s_\gamma(T_{\rm dec})} = \frac{s_\nu(T)}{s_\gamma(T)}\,,
\label{eq:entropy}
\end{align}
where $T_{\rm dec} \sim 2 \rm\ MeV$ is the temperature of neutrino decoupling.  As the Co-SIMPs are ceasing to be relativistic near these energy scales, the effects of their nonzero masses must be taken into account.  The number of relativistic degrees of freedom is defined by  
\begin{align}
\frac{\rho_{\rm tot}(T)}{\rho_{\gamma}(T)} = 1+ \frac{7}{8} \left( \frac{4}{11}\right)^{4/3}  N_{\rm eff}(T) \,.
\label{eq:Ndef}
\end{align}
Solving these two equations gives $N_{\rm eff}(T)$.

\begin{figure}[t!]
\includegraphics[width=0.40\textwidth]{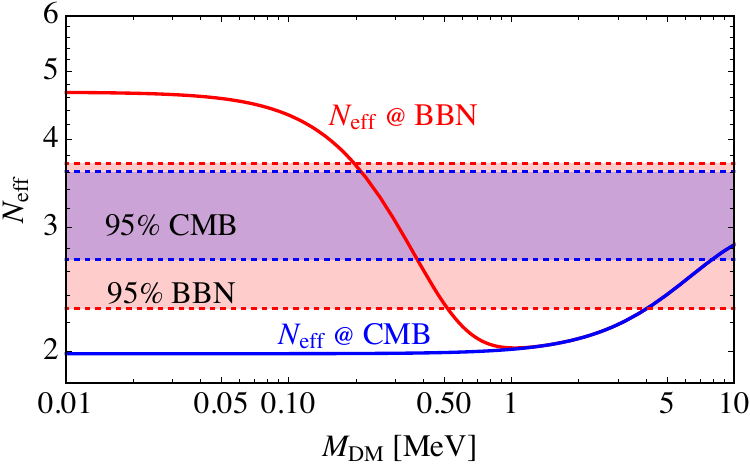}
\caption{Total $N_{\rm eff}$ in the leptophilic Co-SIMP scenario as a function of Co-SIMP mass, with one curve for the BBN epoch and the other for the CMB epoch.  The red and blue bands indicate the $95\%$ confidence intervals from BBN~\cite{Blinov:2019gcj} and CMB~\cite{Aghanim:2018eyx} observations.  
}
\label{fig:NeffAlone} 
\end{figure}

Figure~\ref{fig:NeffAlone} shows $N_{\rm eff}$ in the Co-SIMP scenario (with no other new particles) as function of mass, at the temperatures of the BBN and CMB epochs.  Surprisingly, the Co-SIMP contribution amounts to a $\Delta N_{\rm eff}  \approx - 1$ in the CMB epoch.  This is an interesting counterpoint to many new-physics models that have positive $ \Delta N_{\rm eff}$, e.g., Refs.~\cite{0808.3137, 1006.5276, 1305.1971, 1906.01739}.  The Co-SIMP mechanism is different from a low-scale reheating scenario, where neutrinos are not fully thermalized~\cite{astro-ph/0403323, 1511.00672, 1908.10189}.  Instead, here additional heating of the photon bath leads to a lower $N_{\rm eff}$.  Without additional degrees of freedom, the Co-SIMP mass would have to be above several MeV.

\begin{figure}[t!]
\includegraphics[width=0.40\textwidth]{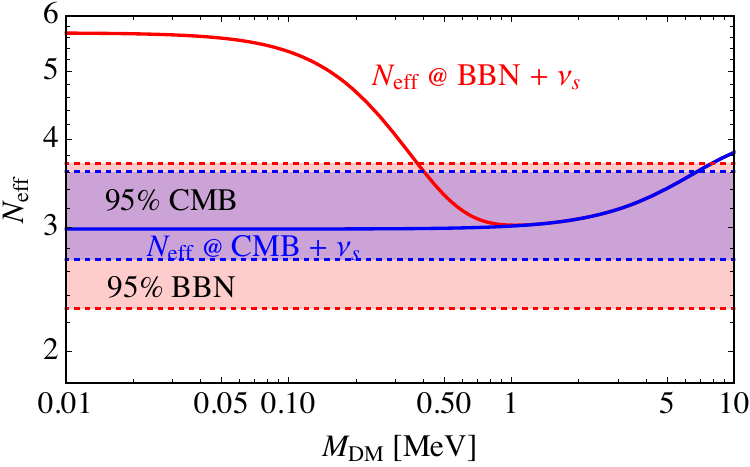}
\caption{Same as Fig.~\ref{fig:NeffAlone}, but where there is one sterile neutrino that equilibrates early~\cite{Abazajian:2012ys}.}
\label{fig:NeffNu} 
\end{figure}

\begin{figure}[t!]
\includegraphics[width=0.40\textwidth]{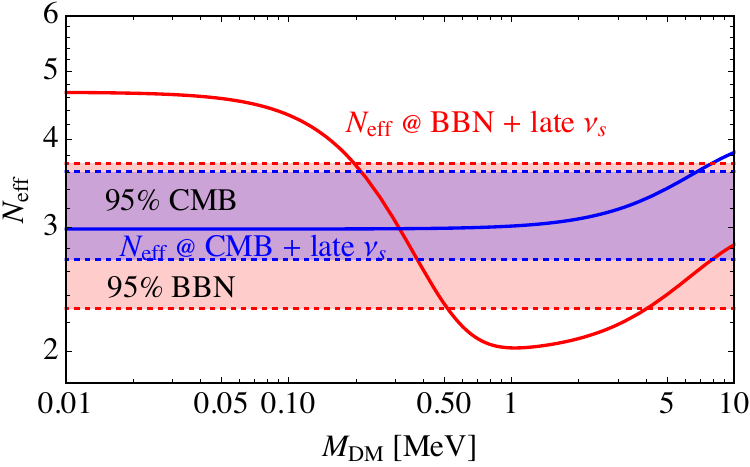}
\caption{Same as Fig.~\ref{fig:NeffAlone}, but where there is one sterile neutrino that equilibrates late~\cite{Kreisch:2019yzn}.}
\label{fig:NeffNuLate} 
\end{figure}

However, with minimal extensions, lower masses are allowed.  Figure~\ref{fig:NeffNu} shows $N_{\rm eff}$ in the Co-SIMP scenario with one additional sterile neutrino that is in equilibrium early~\cite{Abazajian:2012ys}.  Then $N_{\rm eff}$ is compatible with observations in the BBN and CMB epochs if the Co-SIMP mass is larger than about 0.4 MeV.  Alternately, Fig.~\ref{fig:NeffNuLate} is for the case where the sterile neutrino comes into equilibrium after neutrino decoupling~\cite{Kreisch:2019yzn}.  Then the Co-SIMP mass bound is relaxed to about 0.2 MeV.  A Co-SIMP mass below $0.2 \,\rm MeV$ is only compatible with the BBN predictions if the $T_\nu$ in Eqn.~(\ref{eq:entropy}) is lowered with respect to its standard model value. This can occur if a dark sector equilibrates with the neutrino bath, as discussed in Ref.~\cite{Berlin:2017ftj}. We thus argue that laboratory tests of the Co-SIMP mechanism and the determination of its mass can provide new insights into the physics of light degrees of freedom in the early universe.

\subsection{XENON1T Excess}

Very recently, the XENON1T collaboration reported a 3.2-$\sigma$ excess of low-energy events in their electron-recoil spectrum~\cite{Aprile:2020tmw}. The excess could be generated by an unaccounted-for tritium contamination, as the collaboration has argued, and that hypothesis may be testable in the near future~\cite{Robinson:2020gfu}. On the other hand, many beyond-standard model scenarios have been proposed after the observation of the excess~\cite{Baryakhtar:2020rwy, Nakayama:2020ikz, Khan:2020vaf, Cao:2020bwd, Lee:2020wmh, Paz:2020pbc, AristizabalSierra:2020edu, Choi:2020udy, Buch:2020mrg, Dey:2020sai, Chen:2020gcl, Bell:2020bes, DiLuzio:2020jjp, Harigaya:2020ckz, Bally:2020yid, Boehm:2020ltd, Amaral:2020tga, Fornal:2020npv, Alonso-Alvarez:2020cdv, Kannike:2020agf,OHare:2020wum,Zu:2020idx,Lindner:2020kko,Gao:2020wer,Bramante:2020zos,Bloch:2020uzh}.

\begin{figure}[t!]
\includegraphics[width=0.40\textwidth]{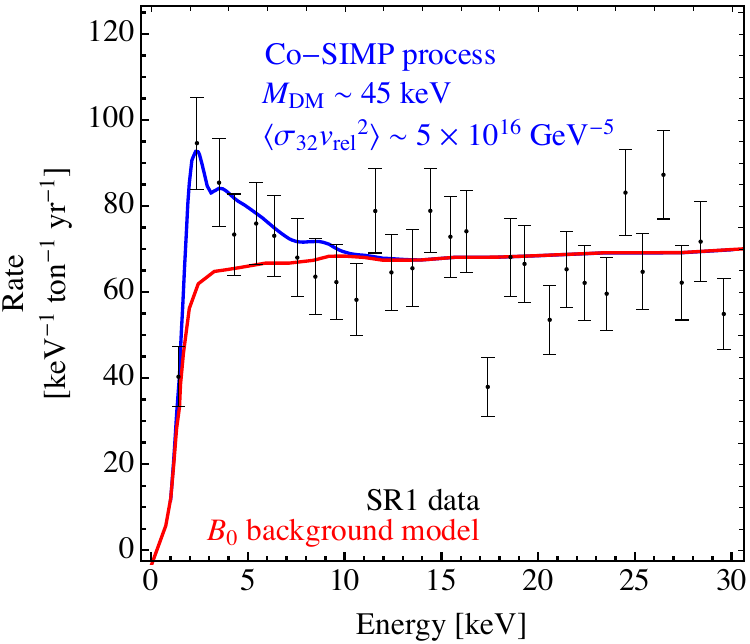}
\caption{Excess of electron-recoil events observed by the XENON1T experiment. The red line shows the expected background, while the blue line shows the background plus a Co-SIMP signal.  The fitted mass and rate factor are found to be on the relic-density line.}
\label{fig:XENON1T} 
\end{figure}

Here we show that the leptophilic Co-SIMP signals we predicted in the main text, before announcement of the XENON1T results, can explain the observations well.  Figure~\ref{fig:XENON1T} shows the resulting recoil spectrum.  From an approximate fit, we obtain $m_{\chi } \approx 45 \, \rm keV$ and $\langle \sigma_{32} v_{\rm rel.}^2 \rangle \approx 5\, \times 10^{16} \, \text{GeV}^{-5}$.  Importantly, this is on the line predicted by the relic abundance computation even though that constraint was not applied in the fit.  Here, instead of treating the signal as intrinsically monoenergetic but subject to detector energy resolution, we also take into account the momenta of the electrons on the orbitals of the xenon atom, which smears the signal somewhat from a line at $\sim 5$ keV (this is well below 45 keV because of the non-linear relation between the Co-SIMP mass and the recoil energy, given by the kinematics in Eg. \ref{eq:recoilenergy}).  For small masses, taking into account the atomic orbitals also slightly changes the excluded regions shown in Fig.~\ref{fig:DDconstraint}, so that these parameters are now allowed (see Fig.~\ref{fig:constraintsComparison}). 

The shape of the signal varies with the target due to how bound the atomic electrons are.  Thus for argon, which has a lower atomic number, our predicted spectrum would be more peaked around 5 keV.  In the DarkSide experiment, which uses argon, there is a low-energy excess of electron recoils observed~\cite{Agnes:2018oej}, though it is accompanied by an excess of nuclear recoils~\cite{Agnes:2018ves}.  We reserve further consideration for future work.

\subsection{Gamma-Ray Production by Co-SIMPs}
 The leptophilic process $\chi  + \chi + e \rightarrow \chi + e$ in Fig.~\ref{fig:CoSIMP}  produces recoil electrons that can produce gamma rays in several ways. The energetic electrons can emit bremsstrahlung, either as final-state radiation (internal bremsstrahlung), or subsequently when scattering off atomic nuclei in the surrounding medium (external bremsstrahlung). We focus on another process, which leads to a more specific signature. The monoenergetic electron recoils can eject electrons from atomic orbitals, which then will be populated by cascade reactions. In the atomic de-excitation processes, monochromatic photons will be emitted. Hence those processes lead to a sharp edge in the photon spectrum, which is model specific, as it encodes the Co-SIMP mass. 
 
 Assuming that a substantial fraction of electrons is bound in low-lying orbitals, the Co-SIMP ejection process will often lead to cascade reactions with monochromatic photons. We thus estimate the resulting photon flux at Earth to be
\begin{align}
& \frac{d \phi}{ d E_\nu} = \frac{1}{8 \pi} \frac{\langle \sigma_{32} v_{\rm rel}^2 \rangle}{M_{\rm DM}^2 M_{\rm SM}} \, J_{32} \, \frac{dN }{dE_\nu}\,, \\ \nonumber
& \text{where } J_{32} = \int_{\Delta \Omega } d \Omega ' \int_{\rm l.o.s.} d \ell \, \rho_{\rm DM}^2 \rho_{\rm SM}\,.     
\end{align}
In contrast to the usual case, here the $\rho_{\rm SM}$ factor makes the signal vary substantially depending on the astrophysical environment. The resulting J-factor for a typical dwarf galaxy is $\sim10^{18}\text{ GeV}^3\text{cm}^{-8}$, which results in a flux below $\sim 10^{-15}\,\text{cm}^{-2}\,\text{s}^{-1} $, too small to be detected.

An intriguing opportunity arises in the case of the galactic center (GC), where the peaked DM and SM matter densities, in particular around the compact SgA* object~\cite{astro-ph/0304125}, lead to an enhanced J-factor $\sim10^{26}\text{ GeV}^3\text{cm}^{-8}$. Low DM masses are favored due to the high number density. The resulting signal is an x-ray flux $\sim 10^{-6}\,\text{cm}^{-2}\,\text{s}^{-1} $ at keV energies. The analysis of $\sim 3$ ks of Chandra x-ray data~\cite{astro-ph/0102151}, at short exposure, does not constrain the thermally predicted cross sections. However, planned x-ray missions, such as Athena~\cite{1807.06092}, with improved energy resolution and effective area, can test relevant parameter space.
 
There is also a loop-induced semi-annihilation process, which leads to larger photon fluxes, but is less specific. The lepton line of the $\mathcal{O}_\ell$ operator can be closed into a loop, which leads to processes of the type $\chi + \chi \rightarrow \chi + n\, \gamma$. The radiation of one photon is forbidden by spin-statistics, so that the leading process is  $\chi + \chi \rightarrow \chi + 2 \gamma$. In the limit $\MDM \ll M_{\rm SM}$, the cross section is
 \begin{equation}
 E_\gamma \, \frac{d \sigma_{\chi \gamma \gamma}}{d E_\gamma} = \frac{\alpha_{\rm EM}^2 M_{\rm SM}^2  x_\gamma^2 \, \left(\frac{3}{4} - x_\gamma \right) }{16 \pi^5 \Lambda^4 \left(1 -  x_\gamma \right) }\,,
 \end{equation}
 where $x_\gamma = E_\gamma/\MDM$ and $0< x_\gamma < 3/4$ is the kinematically accessible range. Again, using the $3 \rightarrow 2$ cross section, given the electron Co-SIMP coupling operator $\mathcal{O}_\ell$, the values of the $\langle \sigma_{32} v_{\rm rel.}^2\rangle $ cross section can be now directly translated into an expected photon flux from the GC, which is of the order of $\sim 10^{-3}\,\text{cm}^{-2}\,\text{s}^{-1} \,\deg^{-2}$ for a mass of $\sim 100$ keV.
 
The flux bounds of Ref.~\cite{1609.00667} from NuSTAR observation of the GC are not competitive with the direct detection limits. However, the planned Athena experiment, under the assumption of  $\sim 300 \,\text{ks}$ observation time, can explore relevant parameter space. It appears that this signal with a broader spectrum can be effective in constraining the model parameters. In the case of a signal discovery, however, a sharp edge in the spectrum as predicted by the Co-SIMP freezeout process would be needed to provide relevant model information.


\subsection{Co-SIMP Stellar Cooling}
In the hot, dense cores of stars and supernovae, the process $e^{+} + e^{-} \rightarrow \chi + \chi + \chi$ could occur, which would provide a new cooling channel. Because this process requires the presence of positrons, the plasma temperature needs to be of order $m_e$. The relevant systems, therefore, are supernovae and  pre-supernova stars. Note that the situation is different from the models with a light scalar mediator, such as discussed in Refs.~\cite{1611.05852,1709.07882}, where a light scalar can mix with the induced longitudinal mode of the photon in the stellar plasma. Since in the Co-SIMP effective theory this is not possible, the above constraints do not apply. However, plasma mixing might become relevant in UV completions with scalars coupled via a Yukawa interaction to electrons. 

The cooling is mitigated by the opacity of the stellar material to the escaping $\chi$, which is determined by  $\chi + e \rightarrow \chi +\chi +e $. This process has a threshold and is only possible if $\sqrt{s} > \MDM$. To investigate the production efficiency and the escape probability of Co-SIMPs, we have to evaluate those cross sections given the supernova environment parameters. 

We model the forming supernova in a very simplistic way, (for a more detailed model, see Ref.~\cite{SNModel}). The inner part of the protoneutron-star (PNS) is assumed to be a sphere of $R_{\rm pns} \approx 5 \, \rm km$ radius, a temperature of $T_{\rm pns} \approx 50 \, \rm MeV$, and a density of $\rho_{\rm pns} \approx 10^{14} \, \text{g}/\text{cm}^3$. The outer region is assumed to have a depth of $R_{\rm out} \approx 100 \, \rm km$, a temperature of $T_{\rm pns} \approx 0.1 \, \rm MeV$, and a density of $\rho_{\rm out.} \approx 10^{9} \, \text{g}/\text{cm}^3$.

The semi-elastic scattering cross section, at collision energies above the electron mass, is $\sigma_{ \chi + e \rightarrow  \chi +  \chi + e} \approx \MDM \sqrt{s}/(256 \pi^3 \Lambda^4) \approx 10^{-28} \left(T/\MDM\right) \text{ cm}^2$. 
Given the PNS temperature, this results in $\sigma_{ \chi + e  \rightarrow  \chi + \chi + e} \approx 10^{-28} (0.1\,\text{MeV}/\MDM) \text{ cm}^2$, leading to a mean free path in the outer region of $  \lambda_{\rm out.}^{-1} \approx \langle \sigma_{ e^+ \chi} v_{\rm rel} \rangle  \rho_{\rm out.}/m_p < 2 \, (\MDM/ (0.1 \text{ MeV}))\text{ m}$. Therefore, any produced Co-SIMP is trapped inside the forming supernova and the bound of Ref.~\cite{SNrefs} does not apply. 

In a pre-supernova star, the temperature rises gradually towards $m_e$. In the edge case ($\MDM \sim m_e$), Co-SIMPs might be produced at the threshold of the semi-elastic scattering reaction, such that some of them escape the star. In this case, even their velocity distribution will affect their mean free path. A detailed study of this effect, however, is beyond our scope.


\subsection{Co-SIMP Self-Scattering}
We investigated whether Co-SIMP processes can affect galactic structure by energy transport or injection. Even though the fundamental interaction cross section is strong, the probability of a Milky Way particle interacting is negligible, due to the low $\rho_{\rm SM}$ of the interstellar medium. However, there is another possibility.

\begin{figure}[t!]
\includegraphics[width=0.40\textwidth]{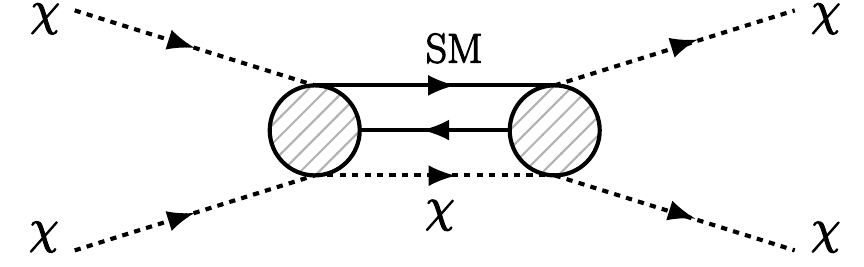}
\caption{Elastic dark matter self-scattering process induced by two-loop topology.}
\label{fig:self} 
\end{figure}

Figure~\ref{fig:self} shows the induced topology for Co-SIMP elastic self scattering:  $  \chi  +    \chi \rightarrow  \chi  + \chi$. The computation of the self scattering cross section leads to $\sigma_{\rm self} \approx 10^{-37}(\text{MeV} /\MDM)^2 \left[ \langle \sigma_{32} v_{\rm rel.}^2 \rangle / (10^{14} \text{ GeV}^{-5})\right] \text{ cm}^2$ in the leptophilic scenario and $\sigma_{\rm self} \approx 10^{-27}(\text{MeV} /\MDM)^2 \left[ \langle \sigma_{32} v_{\rm rel.}^2 \rangle / (10^{14} \text{ GeV}^{-5})\right] \text{ cm}^2$ in the nucleophilic scenario. This difference can be understood from the point of view of the effective interactions. The loop-induced operator for the self-scattering process violates the $Z_3$ symmetry of the dark sector. This violation is connected to the mass of the SM particle and vanishes in the chiral limit. Since the nucleon masses are significantly larger than the electron mass, the self-scattering loop is larger. 

The limit inferred from the Bullet Cluster observations is $\sigma_{\rm self}/M_{\rm DM} <  \text{cm}^2\text{ g}^{-1} \approx 10^{-27} \text{cm}^2\text{ MeV}^{-1} $~\cite{1608.08630}.
In the leptophilic scenario, it becomes only relevant for DM masses around $\sim 10$ keV, which are already excluded by direct detection.  In the nucleophilic scenario, however, it becomes relevant at MeV masses, thus providing the strongest constraints on the low mass end of the nucleophilic Co-SIMP scenario. 

Note that Co-SIMP interactions depositing energy in compact objects cannot test this scenario. In the mass range we consider, the Co-SIMPs would evaporate from bound orbits inside the Sun or the Earth, and the bounds from Ref.~\cite{0705.4298} do not apply. 

\subsection{Low Threshold Detector Sensitivities}

The Co-SIMP process can be tested directly in neutrino detectors, however, at masses above the MeV scale, the even rates quickly drop and detection becomes challenging. At the same time, elastic interactions with SM particles are induced at the two-loop level, and given the large Co-SIMP number densities, event rates are expected to be considerable. We demonstrate that the development of a detector with a low energy threshold will probe relevant Co-SIMP parameter space, even with small exposures. 

\begin{figure}[t]
\includegraphics[width=0.48\textwidth]{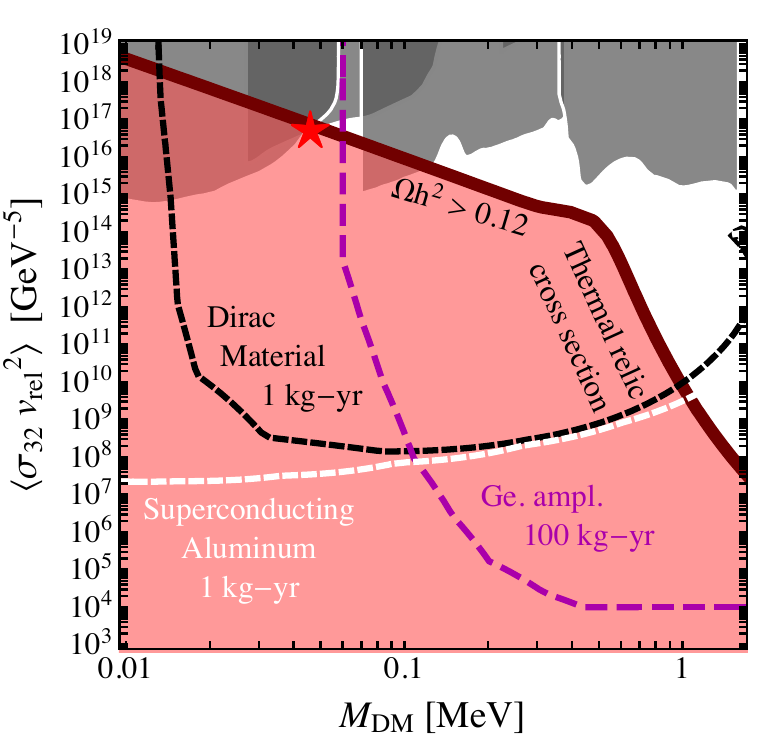}
\caption{Current bounds (gray shaded) and projected sensitivities to electrophilic Co-SIMPs. We show the estimated sensitivities from a detector based on a Dirac material (black dashed line), a superconducting aluminum detector (white dashed line), and germanium crystal with low-energy signal amplification (magenta dashed line). Compared to Fig.~\ref{fig:DDconstraint}, here we take into account the electron-binding effects for xenon, which slightly weaken the limits for low Co-SIMP masses, allowing the Co-SIMP model to explain the XENON1T excess for parameters marked with the star.
}
\label{fig:constraintsComparison} 
\end{figure}

Figure~\ref{fig:constraintsComparison} shows the estimated sensitivities from a detector based on a Dirac material (with 1 kg-yr), a superconducting aluminum detector (with 1 kg-yr), and a low-threshold germanium detector (with 100 kg-yr).  Any of the new technologies for low threshold detectors can test large fractions of the thermal parameter space. It seems promising that the combination of the different detection techniques could test this scenario entirely. 

Note that in all cases, only the neutrino background has been considered~\cite{1512.04533,1604.08206,1708.08929}. However, since the event rates at the low DM masses are so high, even greater realism should lead to a powerful sensitivity reach. 


\subsection{Co-SIMPs and SM Bound States}
In a setup where the Co-SIMPs couple to right handed leptons only, the interaction will be spin sensitive, similarly to as in Ref.~\cite{1401.6457}. The two-loop topology shown in Fig.~\ref{fig:self} induces a four-lepton interaction that affects the hyperfine splitting of leptonic bound states. A determining factor for the sensitivity of a system to this short-range force is the compactness of the wave function at the origin, which scales as the inverse Bohr radius cubed, $\sim (\alpha_{\rm EM} m_\ell)^3$.  We estimate the energy shift to be
\begin{equation}
\Delta E^{\ell^+ \ell^-}_{\rm hfs.} \approx - \frac{\alpha_{\rm EM}^3 m_\ell^3}{2^{11} \pi^5\, \Lambda^2} \log{\left( \frac{\Lambda^2 + \MDM^2}{4 \MDM^2} \right)}\,.
\end{equation}

For positronium, the energy difference between the spin-singlet and triplet configurations is $\Delta E^{e^+ e^-}_{\rm hfs.} \approx - 43 \left(\text{GeV}/\Lambda \right)^2\, \rm Hz$. This is only an order $\sim 10^{-10}$ correction, well below the experimental sensitivity~\cite{1310.6923}. However, assuming universal leptonic coupling strength, the effect is much stronger for a  muon-anti-muon bound state, which has a much more compact wave-function, so $\Delta E^{\mu^+ \mu^-}_{\rm hfs.} \approx - 380 \left(\text{GeV}/\Lambda \right)^2\, \rm MHz$. A measurement of the hyperfine splitting in true muonium with $\sim 10\, \%$ precision would probe thermally produced Co-SIMPs with masses below $\sim 30 \rm \,keV$. With a $\sim 1\, \%$ precision, masses below $\sim 0.3 \rm \,MeV$ would be tested, covering most of the relevant parameter space. Direct probes of this exotic atom are conceivable in the near future, see for example Ref.~\cite{1904.08458}.

\subsection{Possible UV Completions}
We briefly sketch two ideas for renormalizable Co-SIMP models. 

In the leptophilic scenario, two new ingredients are needed. First, a vector-like fermion that couples through a Yukawa interaction to the Higgs field and the lepton doublet.  Second, a scalar field that has a Yukawa interaction with the new fermion and the right-handed SM lepton field, and which couples via a quartic interaction to the Co-SIMP, respecting the $Z_3$ symmetry.  After electroweak symmetry is broken and the new fermion and scalar are integrated out, the resulting low energy operator is  $\mathcal{O}_\ell$, which we have used for the relic density calculations. 

In the nucleophilic case, a new sector with strong dynamics can induce a Wess-Zumino type interaction among five dark-sector pions, similarly to~\cite{NNLOSIMPs,SIMPlest,ThermalHistory}. Two dark pion species can be made heavy and with appropriate quantum numbers for their constituent quarks, a mixing with SM pions could be generated. Integrating out the two heavy dark pions induces an $\mathcal{O}_\pi$ operator at low energies. Alternatively, a dilaton type scalar can be coupled to the dark sector pions. The dilaton-Higgs mixing would result in multiple operators coupling the dark sector to the SM, the size of the coupling would be determined by the size of the Higgs Yukawa couplings, favoring a top-philic model.

Those are only two possible implementations, and more scenarios are conceivable, for example number changing processes based on spin-2 field interactions~\cite{1708.06764}. 


\clearpage

\footnotesize
\bibliographystyle{abbrv}



\end{document}